\documentstyle[twoside,fleqn,espcrc2,epsfig]{article}

\newcommand{\AmS}{{\protect\the\textfont2
  A\kern-.1667em\lower.5ex\hbox{M}\kern-.125emS}}

\title{A Polarized HERA Collider}

\author{V. W. Hughes\thanks{Presented by V. W. Hughes; This work was supported
        in part by the US Department of Energy.}\address{Physics
        Department, Yale University
        P.O.Box 208121, New Haven, CT 06520-8121, U.S.A.}
        and
        A. Deshpande$^{\rm a}$
        }

\begin{document}

\begin{abstract}
A brief review is given of the status of nucleon spin structure functions as determined
from polarized deep inelastic lepton-nucleon scattering, including current
outstanding problems.  The characteristics of a polarized HERA
collider, some of the particle physics topics it could address, and the
accelerator physics challenges it must meet are discussed.
\end{abstract}

% typeset front matter (including abstract)
\maketitle

\section{Introduction}
It is often said that physics studies involving spin can lead to surprising
results.  We list some outstanding examples below.

\begin{center}{\bf Some Surprises With Spin}
\end{center}

1. Space quantization associated with quantized spin directions. Stern, Gerlach,
1921.

2. Atomic fine structure and electron spin magnetic moment.  Goudsmit, Uhlenbeck,
 1926.

3. Proton anomalous magnetic moment;  \\$\mu_{p}=2.79~{\rm nm}$.  Stern, 1933.

4.  Electron spin anomalous magnetic moment.  $\mu_{e} = \mu_0$ (1.00119); QED.
Kusch, 1947.

5.  Electroweak interference from $\vec{e}_{1}d$ DIS parity nonconservation. 
Prescott \& SLAC-Yale Collaboration, 1978.

6.  Proton spin structure;  puzzle or crisis.  EMC, 1989.\\
Hence from an historical viewpoint the spin variable is a promising one for
discovery.

The first experiment, beginning in the mid-1970's, on polarized lepton-proton deep inelastic scattering was a
series of two measurements at SLAC by a SLAC-Yale group, using an atomic beam
polarized electron source built at Yale and an electron beam energy of $\sim$20~GeV. 
The virtual photon-proton asymmetry
$A^{\rm p}_{1}$
was measured in the $x$ range from 0.1 to 0.7 and found to be large.  The 
values
were consistent with a plausible quark-parton model satisfying the Bjorken sum
rule, and also satisfying the Ellis-Jaffe sum rule if Regge extrapolation of
$g^{\rm p}_{1}(x)$ to $x$ = 0 was employed~\cite{Yslac}.

The European Muon Collaboration (EMC) in the mid-1980's made a similar
measurement of $A^{\rm p}_{1}$ using a polarized $\mu^+$ beam with energy up to
200~GeV.  In addition to confirming the SLAC results for $x>0.1$, their data 
extended
down to $x=0.01$.  Unexpectedly, the  $A^{\rm p}_{1}$ data at low $x$ fell 
well below
the extrapolated SLAC data.  The consequence was violation of the Ellis-Jaffe
sum rule at the 3 standard deviation level, and the conclusions were 
that only the
small fraction $\Delta=0.12\pm 0.17$ of the proton spin is due to quark spins
and that strange quarks have a substantial negative polarization~\cite{EMC}.

This discovery by EMC has led to major new experiments at CERN, SLAC and DESY
which were reviewed by R. Windmolders in this workshop.

Some outstanding questions relevant to spin structure remain:  1) Behavior of
$g^{\rm p}_{1}(x)$ at small $x$ and the value of the first moment 
$\Gamma^{\rm p}_{1}$=$\int^{1}_{0}dx$$g^{\rm p}_{1}(x)$.  2) Contribution 
of gluons to proton spin and
the polarized gluon distribution. 3) Contribution of the orbital angular 
momentum
of quarks and gluons to proton spin. 4) Hadronic spin structure of the photon.
5) Chiral structure of any observed contact interaction or leptoquark beyond
the standard model.

\begin{table*}[hbt]
% space before first and after last column: 1.5pc
% space between columns: 3.0pc (twice the above)
\setlength{\tabcolsep}{1.5pc}
% -----------------------------------------------------
% adapted from TeX book, p. 241
\newlength{\digitwidth} \settowidth{\digitwidth}{\rm 0}
\catcode`?=\active \def?{\kern\digitwidth}
% -----------------------------------------------------
\caption{A Polarized HERA Collider}
\label{tab:polhera}
%\begin{tabular*}{\textwidth}{@{}l@{\extracolsep{\fill}}rrrr}
%\hline
%                 & \multicolumn{2}{l}{Electron Beam}
%                 & \multicolumn{2}{l}{Proton Beam} \\
\begin{tabular*}{\textwidth}{@{}l@{\extracolsep{\fill}}rr}
\hline
                 & Electron Beam
                 & Proton Beam \\                 
%\cline{2-3} \cline{4-5}
%                 & \multicolumn{1}{r}{Energy}
%                 & \multicolumn{1}{r}{Polarization}
%                 & \multicolumn{1}{r}{Energy}
%                 & \multicolumn{1}{r}{Polarization}     \\
\hline
Beam Energy              & $26-30$ GeV & $800-930$ GeV \\
Polarization Status      & Polarized Sokolov  & Negligible polarization \\
                         & Ternov Effect (SKE) & due to SKE  \\ 
                         & P$_{e}$ $\sim 60\%$ in 1/2 hour  & Need polarized source \\
Expected Polarization    & 70\% & 70\%\\
Uncertainty $\delta P/P$ & $\le 2\%$ & $\le 3\%$ \\
Integrated Luminosity    & \multicolumn{2}{r}{$\sim 500$ pb$^{-1}$, 3 years 
running with 150--170 pb$^{-1}$/year} \\
\hline
\end{tabular*}
\end{table*}

Experiments such as polarized lepton-nucleon scattering which require a
polarized beam and a polarized target are sometimes spoken of as spin physics
and are sometimes regarded as not of central interest. However, it is well 
known that
the hadronic tensor $W^{\mu\nu}$, which describes the proton in DIS $ep$ 
or $\mu p$ scattering, involves four scalar functions: $F_{1}(x)$, $F_{2}(x)$,
$g_{1}(x)$ and $g_{2}(x)$, which are required for a complete knowledge of 
$W^{\mu\nu}$.  The functions $F_{1}$ and $F_{2}$ do not involve the spin
variables, whereas $g_{1}$ and $g_{2}$ do.  Perhaps the principal achievements
of HERA to date are the measurement of $F_{2}(x)$ at small $x$ and also at high
$Q^{2}$, the determination of the unpolarized gluon distribution in the
proton, study of the hadronic constituents of the photon, and extension of 
the limits on a
contact interaction or leptoquark search.  A polarized HERA collider will
address these same topics from a different viewpoint. Spin is a fascinating
tool but it is not the goal of these experiments.

\section{Characteristics of a Polarized HERA Collider}
There has been considerable interest in a possible polarized
HERA collider with the characteristics indicated in Tab.~\ref{tab:polhera},
in several workshops associated with HERA\cite{polhera,ws99}. At present of course there is
a polarized electron beam in HERA which is used in the HERMES experiment.
Development of a polarized proton beam is required. Then one or both of the
existing collider detectors --ZEUS,H1-- could be used for measurements
of polarized $e^{\pm}p$ DIS.

\begin{figure}[here]
\epsfxsize=6.0cm
\hfil
\epsffile[35 35 400 400]{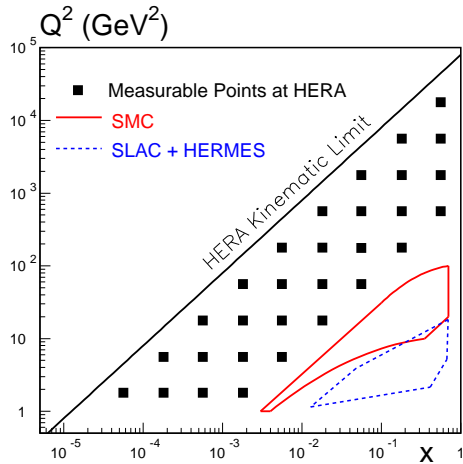}
\hfil
\caption{The $x-Q^{2}$ range of HERA compared to the fixed target 
experiments at CERN, SLAC and DESY.}
\label{fig:xq2}
\end{figure}

The huge increase in the $x-Q^{2}$ range for measurements of spin
variables possible with a polarized HERA collider is shown in Fig.\ref{fig:xq2}.
Two orders of magnitude increase in both $x$ and $Q^{2}$
range is possible for exploring the spin structure of the proton. 
If a sufficiently intense source of $^{3}$He$^-$ can be developed, measurements 
of the spin structure of the neutron could be done\cite{polhera}.

\section{Particle \& Nuclear Physics with Polarized HERA}
\label{sec:physics}
\subsection{Measurement of $g_{1}^{\rm p}$ at low $x$}
\begin{figure}[t]
\epsfxsize=6.5cm
\hfil
\epsffile[20 80 525 540]{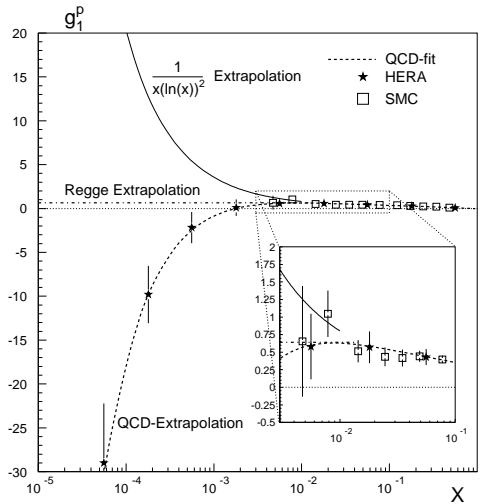}
\hfil
\caption{The statistical 
uncertainty on $g_{1}^{\rm p}$ from possible measurements at HERA with 
500 pb$^{-1}$ is shown along with different theoretical predictions for 
the low $x$ in the kinematic region $x < 0.003$.}
\label{fig:lowx}
\end{figure}

The behavior of $g_{1}^{\rm p}$ at low $x$ is of fundamental interest
and the largest uncertainty on the first moment of $g_{1}^{\rm p}$ now
comes from the unmeasured low $x$ region, $x < 0.003$~\cite{SMC}. 
The statistical
uncertainties for a measurement of $g_{1}^{\rm p}$ with a polarized HERA 
collider (Tab.~\ref{tab:polhera}), using the H1 or ZEUS detector, 
are shown in Fig.~\ref{fig:lowx}.
Even though the predicted asymmetries are as small as $3\times 10^{-4}$ at
low $x$, false asymmetries associated with correlations of proton beam
intensity or bunch crossing angle with proton polarization, or with 
time variation of detection efficiencies can be kept still smaller by
modulation of the spin direction. From the polarized
H$^{-}$ source any desired polarization can be provided for a proton bunch 
so that rapid modulation for successive interactions is achieved. Also 
in the HERA ring
at high energy, all of the spin directions of the proton bunches can be flipped
periodically by a microwave pulse.
Recently~\cite{ws99} it has been checked using a fast
detector simulation that the detector 
smearing and event migration does not hinder the measurement $g_{1}$.
Also shown in Fig.\ref{fig:lowx} are the various 
low $x$ predictions for the structure function arising from models and QCD fits
consistent with the presently available data.

\subsection{The Polarized Gluon Distribution}
Determination of the polarized gluon distribution $\Delta G(x,Q^{2})$
inside a nucleon and its first moment $\Delta G$ have become important 
goals of all experiments proposed
and planned in the next decade. Polarized HERA can contribute significantly, 
through the various different and independent ways
in which it can measure $\Delta G(x,Q^{2})$. 

Perturbative QCD analysis of $g_{1}^{\rm p}$ allows a determination of
$\Delta G(x,Q^{2})$ through a next-to-leading order analysis\cite{SMC}.
The published value of 
$\Delta G = 1.0^{+1.2}_{-0.3}({\rm stat}) ^{+0.4}_{-0.2}({\rm syst}) ^{+1.4}_{-0.5}({\rm theo})$ \cite{SMC}
indicates that statistical and theoretical uncertainties dominate our
lack of knowledge of $\Delta G(x,Q^{2})$. 
A study~\cite{polhera}
using simulated HERA data and the presently available fixed target data
indicates that the statistical \& theoretical uncertainties could be 
reduced by factors $\sim 2$ \& $\sim 3$, respectively. 

\begin{figure}[t]
\epsfxsize=7.0cm
\epsffile[15 320 530 545]{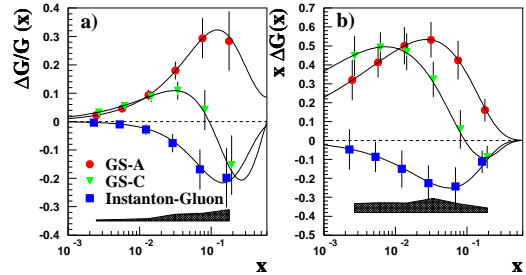}
\caption{Statistical accuracy possible for the measurement of $\Delta G(x)$ 
using 2 jets from the PGF process shown
with different predictions for $\Delta G(x,Q^{2}=20~{\rm GeV}^{2})$.}
\label{fig:dijet}
\end{figure}

In the Photon Gluon Fusion (PGF) process the gluon is involved at leading 
order (LO).
The measurement of such a process 
through detection of 2 high--p$_{T}$ jets and the scattered electron 
allows access to the polarized gluon distribution\cite{polhera}. 
Figure~\ref{fig:dijet} shows the statistical
uncertainty achieved by polarized HERA collider (Tab.~\ref{tab:polhera}).
Also shown are
three widely different predictions for $\Delta G(x,Q^{2})$ at LO, all 
consistent with the fixed target data, which HERA data can easily distinguish between. 
Expected uncertainty on $\Delta G / G$ after HERA measurements of 2-jets is
$\pm 0.1$\cite{polhera}. It has also been checked that the detector smearing 
and migration effects due to the measurement process do not affect the measurability
of $\Delta G$. Recently, NLO corrections to this process were evaluated and found
to be small.

Both H1 and ZEUS collaborations have published results on the
parton distributions inside the unpolarized photon $q^{\gamma}$.
With polarized HERA one could investigate the structure of
the polarized photon $\Delta q^{\gamma}$.
A study\cite{polhera}, using single and 2 high-p$_{T}$ jets or
hadron tracks from the PGF process in photoproduction, showed that 
a luminosity of 100 pb$^{-1}$ was sufficient to resolve the polarized 
photon and the gluon structure.

\section{Polarized HERA: Accelerator Aspects}
\label{sec:machine}
\begin{figure}
\begin{center}
\epsfig{file=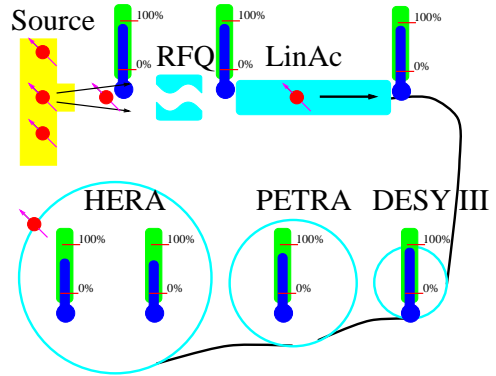,width=6.5cm,angle=-90.0,
        bbllx=15pt,bblly=50pt,bburx=580pt,bbury=780pt}
\end{center}             
\caption{The various stages of the proton acceleration from the polarized
source to the HERA ring.}
\label{fig:chain}
\end{figure}

Production of a high energy polarized proton beam in HERA requires first
a polarized H$^{-}$ source and then acceleration to high energy through the 
DESY acceleration chain with retention of polarization (Fig.~\ref{fig:chain}).
The full initial study\cite{ak} of the requirements for a high energy polarized beam
in HERA identified two crucial problems. One is the development of an adequate
polarized H$^{-}$ source, and the second is the acceleration
and retention of proton polarization in HERA.

\subsection{Polarized Source for HERA}
\begin{figure}
\epsfig{file=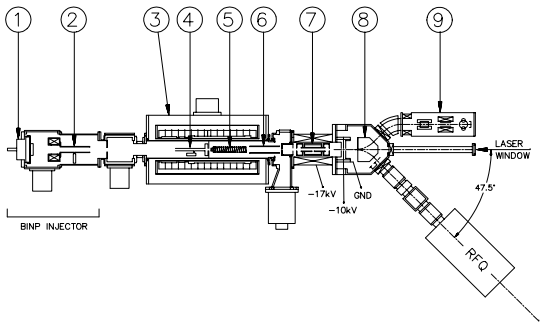,width=7.0cm,angle=90.0,
        bbllx=200pt,bblly=50pt,bburx=530pt,bbury=680pt}
\caption{The TRIUMF Optically Pumped Polarized Ion Source: 1. plasmatron proton
         source, 2. hydrogen neutralizer, 3. superconducting solenoid, 
         4. helium ionizer, 5. optically pumped Rb vapor cell, 6. deflection
         plates, 7. Na vapor ionizer, 8. bending magnet, 9. Lyman-alpha polarimeter
         }
\label{fig:oppis}        
\end{figure}

The design specifications for a polarized H$^{-}$ source for HERA are\cite{ak}: I=20 mA,
in 100$\mu s$ pulses at 0.25 Hz with emittance $2\pi$ mm$\cdot$mr, and $P_{p}=0.8$. At present the most promising approach
is the Optically Pumped Polarized Ion Source (OPPIS) development at TRIUMF
by A. Zelenski {\em et al.}~\cite{oppis}. The overall source arrangement is shown in
Fig.~\ref{fig:oppis}. 

A polarized H$^{-}$ source is now being developed at TRIUMF for the Relativistic
Heavy Ion Collider (RHIC) at BNL for their RHIC-Spin program. 
%The intensity
%requirement is about a factor of 10 less than needed at HERA. However,
Development of an OPPIS source suitable for HERA is progressing with optimism.  

\subsection{Acceleration of Polarized Protons}
\begin{figure}
\epsfig{file=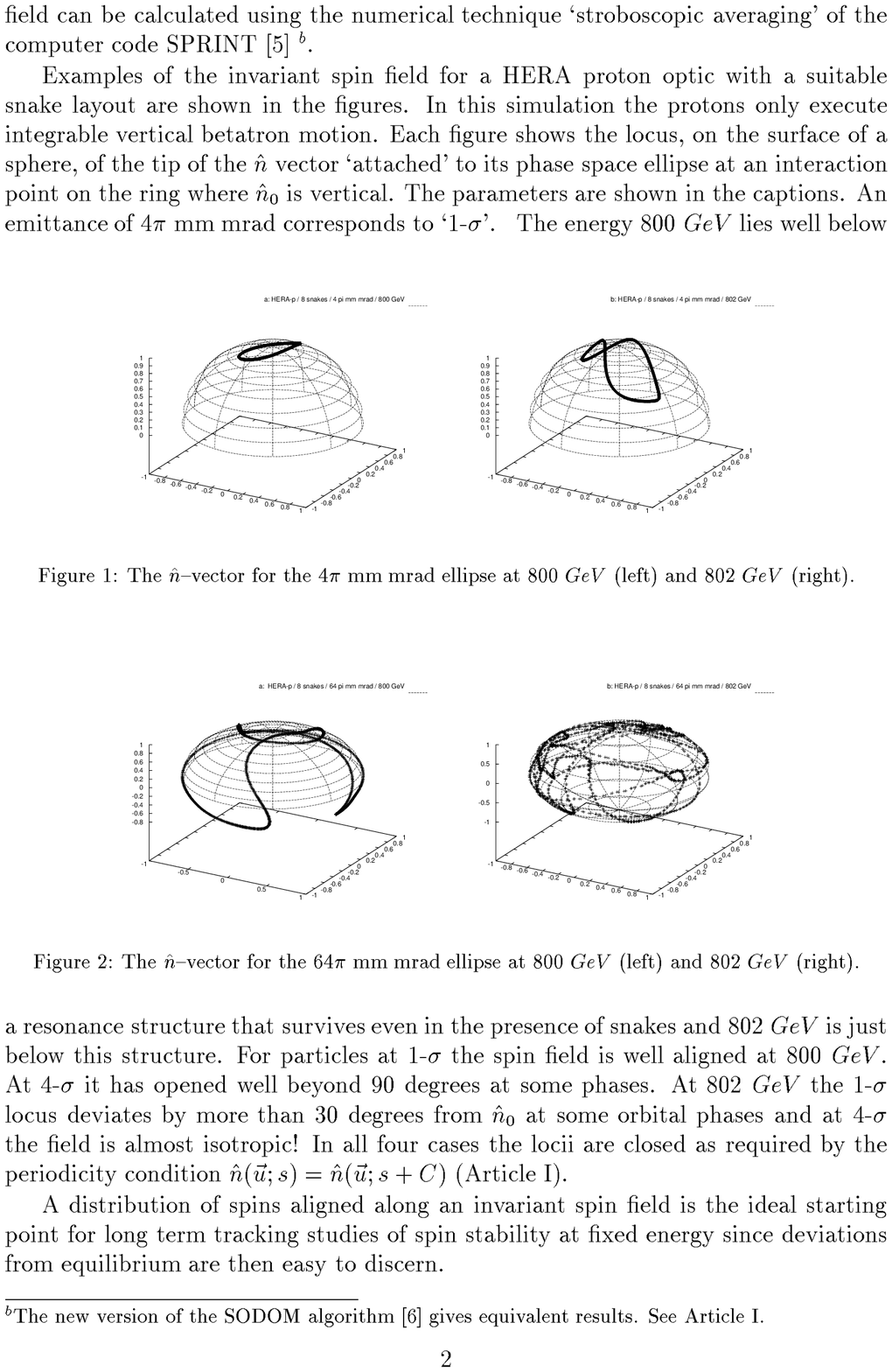,width=6.0cm,clip=,
       bbllx=140pt,bblly=490pt,bburx=290pt,bbury=590pt}\\
\epsfig{file=2figs.ps,width=6.0cm,clip=,
        bbllx=310pt,bblly=490pt,bburx=460pt,bbury=590pt}        
\caption{SPRINT simulation of the resultant spin vector at 800 (802) GeV 
         proton beam above (below).
         Shown is the effect of 4$\pi$ mm mr deviation from the nominal spin direction.}       
\label{fig:sprint}
\end{figure}

The acceleration and then storage of polarized protons from the ion source
to the high energy HERA ring is a major problem dominated principally by 
depolarizing resonances. Because of the large anomalous magnetic moment of the
proton ($\mu_{\rm p}= 2.79~{\rm nm}$), the relativistic equations of spin motion
\cite{barber} show that at high energy the number of spin precessions per 
orbit --the spin tune-- is large, indeed equal to $G\gamma$ which is 
$\sim$1530 at E=800 GeV. Hence the spin motion is very sensitive to the 
magnetic field,
and in particular the many intrinsic and imperfection resonances can lead to 
depolarization. Avoidance of depolarization involves the use of the Siberian Snake
principle\cite{derb}. Extensive simulation calculations by spin tracking codes
have been done\cite{barber}. Figure~\ref{fig:sprint} shows 
for two energies the equilibrium polarization distribution or the spin vector for the stored
proton beam in HERA, using 8 snakes.
Although such results are encouraging, much further intensive
study of spin motion is required to assure adequate proton polarization at
high energy. Electron cooling in the DESYIII ring is being considered to 
reduce the beam emittance and thus make it simpler to obtain high proton
polarization.

\subsection{Proton Beam Polarimetry}
Measurement of proton beam polarization at high energy is under active
development at BNL, mainly for
the RHIC Spin program in which polarized proton beams up to 250 GeV will collide.
Three types of
polarimeter are presently considered\cite{ws99}: 1) Inclusive pion production
$ \vec{p} + C \rightarrow \pi^{+} + X$,
2) $\vec{p}+C$ elastic scattering in the Coulomb Nuclear Interference (CNI)
region, capable of $\sim~8\%$ absolute accuracy determined by the theoretical
uncertainty of the hadronic spin flip amplitude, and 3) $\vec{p} + p$ 
elastic scattering involving a polarized jet target. This method equates the asymmetry
$A$ in the elastic scattering $(p_{\rm b} + \vec{p}_{\rm jet})$, for which
the polarization of the jet $P_{\rm jet}$ is known, to the beam 
polarization $P_{\rm b}$ for $(\vec{p}_{\rm b}+ p_{\rm jet})$
scattering for the same scattering kinematics. It is capable of 
$\sim 3\%$ absolute accuracy.

\section{Summary}
Important physics results can be expected from $\vec{e} + \vec{p}$ collisions 
with the polarized HERA collider. 
However, increased effort is needed {\em now} to
solve the challenging accelerator physics problems associated with
achieving high energy polarized protons in the HERA ring.

\end{document}